\documentclass[11pt,reqno]{amsart}

\usepackage{amssymb,amsmath} 
\usepackage[]{graphicx}

\newcommand{\beq}[1]{  \begin{equation} \label{#1} }  
\newcommand{\eeq}{     \end{equation}}  
\newcommand{\bal}[1]{\begin{align} \label{#1} }

\numberwithin{equation}{section}

\newtheorem{thm}{Theorem}
\newtheorem{lem}{Lemma}

\def\bd#1{\mbox{\boldmath$\displaystyle\mathbf{#1}$} }   
\newcommand{\rf}[1]{(\ref{#1})}
\def\tens#1{\mathbb{\,#1}}	   
\def\tr{\operatorname{tr}} 
\def\dd{\operatorname{d}} 
\def\sgn{\operatorname{sgn}} 
\def\diag{\operatorname{diag}} 

\title[Conjugate stress and strain]{Eulerian conjugate stress and strain}

\author{Andrew N. Norris}
\address{Mechanical and Aerospace Engineering, 
	Rutgers University,  Piscataway NJ 08854-8058, USA }
\email{norris@rutgers.edu}


\begin{document} 


\begin{abstract}

New results are presented for the stress conjugate  to  arbitrary Eulerian  strain measures.   The conjugate stress depends on two arbitrary quantities: the strain measure $f({\bd V})$ and the corotational rate defined by the spin ${\bd \Omega}$.  It is shown that for every choice of $f$ there is a unique spin, called the f-spin, which makes the conjugate stress as close as possible to the Cauchy stress.  The f-spin reduces to the logarithmic spin when  the strain measure is the Hencky strain $\ln {\bd V}$.  The formulation and the results emphasize the similarities in form of the Eulerian and Lagrangian  stresses conjugate to the  strains $f({\bd V})$ and $f({\bd U})$, respectively.  Many of the results involve the solution to the equation ${\bd A}{\bd X}-{\bd X}{\bd A}={\bd Y}$, which is presented in a succinct format.

\end{abstract}

\maketitle

\section{Introduction}\label{intro}
The notion of stress and strain are interlinked, regardless of the existence of a strain energy function.  At the most basic level they are related by mechanical  power, 
the rate of work per unit current   volume of material,
\beq{1}
\tr ({\bd \sigma} {\bd D}) = \dot{w} .  
\eeq
Here ${\bd \sigma}$ is the Cauchy stress  and ${\bd D}$ the stretching tensor.    This work-conjugate relation  is independent of any notion of a reference configuration, although it is useful to introduce one.  Let 
${\bd F}$ be the deformation gradient between the current and reference states, and let 
 ${\bd T}$ and  ${\bd E}$ be the stress and strain  associated with the reference state.  ${\bd T}$ and  ${\bd E}$ are   mutually conjugate if they satisfy 
\beq{-10}
 \tr ({\bd T} \dot{\bd E})=\dot{w}\, \det{\bd F} , 
\eeq
where the factor $\det{\bd F}$ arises  from the change in volume between the current and reference descriptions.  
In fact, eq. \rf{-10} is usually taken as the starting point for determining stress.   The choice  of the strain ${\bd E}$ is not unique, but once chosen it fixes the definition of ${\bd T}$ through the work conjugacy \rf{-10}.  
It is strange but true that the same simple connection does not apply to the relation between current or Eulerian strain and the Cauchy stress.  The difficulty is in the definition of strain, say $\bd e$.  What $\bd e$ is such that $\dot{\bd e} = {\bd D}$?
It turns out that this question is incomplete and that we must broaden it and seek the strain for which 
$\stackrel{\circ}{{\bd e}} = {\bd D}$, where  $\stackrel{\circ}{}$ signifies a corotational rate. Actually, the corotational rate itself also has to  be found.  Fortunately, both the strain and the rate have been determined: Xiao et al.   \cite{Xiao98c}
showed that the unique solution is obtained by the Hencky strain $\ln {\bd V}$ in combination with the logarithmic rate.  But we are getting ahead of ourselves.  

It is evident that work-conjugacy is simpler for  reference or Lagrangian stress and strain  than for their counterparts in the current or Eulerian configuration.   Note that the  distinction between Lagrangian and Eulerian is 
made explicit by the polar decomposition ${\bd F}= {\bd R}{\bd U}= {\bd V}{\bd R}$: quantities associated with or defined by ${\bd U}$ and $ {\bd V}$ will be called Lagrangian and Eulerian,  respectively. 

It is instructive to  review work-conjugacy for Lagrangian stress and strain.   
The starting point is the fact that the stretching tensor ${\bd D}$ is the symmetric part of $\dot{\bd F}{\bd F}^{-1}$. 
Let the   strain be chosen, quite generally, as ${\bd E}= f({\bd U})$ where the function $f$ is sufficiently smooth, then eqs. \rf{1} and \rf{-10} imply
\beq{-2}
\tr \big({\bd T} \big[\nabla f({\bd U}) \big] \dot{\bd U}\big)
=   \tr ({\bd \sigma} {\bd D})\, \det{\bd F}. 
\eeq
The gradient  $\nabla f({\bd U})$ is a fourth order tensor function  which will be described later.  At the same time the  kinematic quantities, strain rate
$\dot{\bd U}$ and stretching  ${\bd D}$, may be related quite easily  (see Appendix \ref{appb})
\beq{3051}
\dot{\bd U}  = 2
(  {\bd U}  \boxtimes{\bd I} +{\bd I}\boxtimes{\bd U}  )^{-1} ({\bd U}\boxtimes{\bd U})\, 
{\bd R}^t{\bd D}{\bd R}.
\eeq
Using the independence of  ${\bd D}$, eqs. \rf{-2} and \rf{3051} imply, 
 formally at least, that the  stress conjugate to the Lagrangian strain $f({\bd U})$ is 
\beq{-4}
  {\bd T}  
  =  \big( \nabla f({\bd U})\big)^{-1} \, {\bd T}^{(1)} , 
\eeq 
where ${\bd T}^{(1)}$, sometimes called the Biot stress or the Jaumann stress,  and    ${\bd S}$, the second Piola-Kirchhoff stress tensor, are 
\beq{1-4}
{\bd T}^{(1)}= \frac{1}{2}  
  ({\bd U}\boxtimes{\bd I} +{\bd I}\boxtimes{\bd U}  )\, {\bd S}, 
  \qquad   \qquad
  {\bd S}  =  {\bd F}^{-1} {\bd \sigma}{\bd F}^{-t} \det{\bd F}. 
\eeq 
We have used the symmetry of $\bd T$ and certain commutative properties to express the stress in \rf{-4} as a fourth order tensor acting on ${\bd T}^{(1)}$.  The tensor product notation, $\boxtimes$, explained in the next Section,  is  used throughout  as we find it makes results more transparent.  Equations \rf{-4} and \rf{1-4} embody  work-conjugacy for arbitrary Lagrangian strain $f({\bd U}  )$. 

Although the notation in eq. \rf{-4} might be unfamiliar the result is not, see \cite[eq. (3.5.31)]{Ogden84}. 
The fourth order gradient tensor $\nabla f({\bd U})$  is discussed in detail by  Norris \cite{Norris07b}.  In particular,  it is positive definite, symmetric and invertible for any strain measure function \cite{Hill78}. 
Examples will be presented for the  Seth-Hill strain measure functions, 
 \beq{777}
 f^{(m)}(x) =  m^{-1} (x^m-1).  
 \nonumber 
 \eeq
For instance,  the  stress ${\bd T}^{(m)}$  associated with  $f^{(m)}({\bd U})$ is 
\begin{subequations}\label{-7}
\bal{-7a}
{\bd T}^{(0)}&=  \big( \nabla \ln {\bd U} \big)^{-1} \, {\bd T}^{(1)} 
= \int_0^1 \dd x\, {\bd U}^{x} \boxtimes {\bd U}^{1 - x}\, {\bd T}^{(1)}, 
 \\
{\bd T}^{(\frac13)}&= \frac{1}{3}  
  ({\bd U}^{2/3}\boxtimes{\bd I} +{\bd U}^{1/3}\boxtimes{\bd U}^{1/3}  +{\bd I} \boxtimes{\bd U}^{2/3}   )\,{\bd T}^{(1)} , \label{-7b1}
\\
{\bd T}^{(\frac12)}&= \frac{1}{2}  
  (\sqrt{\bd U} \boxtimes{\bd I} +{\bd I}\boxtimes \sqrt{\bd U} )\,{\bd T}^{(1)}, \label{-7b}
\\
{\bd T}^{(2)}&=  {\bd S},\label{-7d}
  \\
{\bd T}^{(-m)}&= ({\bd U}^{m}\boxtimes{\bd U}^{m})\,{\bd T}^{(m)}.\label{-7e}
\end{align}
\end{subequations}
Some of the conjugate stresses  listed  are well known, e.g. $m=1,2,-2$ \cite{Hill78,Ogden84}, and \rf{-7e} follows from 
\cite[p. 158]{Ogden84}.  Identities \rf{-7a} -  \rf{-7b} and the others will become evident later.  The second identity in \rf{-7a} follows from \cite{Norris07b}. We note that the Piola-Kirchhoff stress is conjugate to  ${\bd E} = \frac12 ({\bd U}^2 - 1)$, the Green strain, which is typically used in applications. 

A principal objective of this paper is to find analogous expressions for the Eulerian stress ${\bd \tau}$ conjugate to the strain ${\bd e} = f({\bd V})$ where ${\bd V}$ is the right stretch tensor and the function $f$ is again arbitrary.  We also require that  the Cauchy stress be included among the Eulerian stresses, just as the Piola-Kirchhoff stress appears naturally for the Green strain.  

Unlike the Lagrangian strains,  $\dot{f}({\bd V})$ is not an objective tensor \cite{Dill06}, and it is known that this strain rate  does not, in general, possess a conjugate stress \cite{Macvean68}.  This difficulty can be avoided by defining conjugacy in terms of  corotational strain rates.  
The  corotational rate of a symmetric second order tensor ${\bd A}(t)$ is  
\beq{-199}
\stackrel{\circ}{{\bd A}} \equiv \dot{\bd A} + {\bd A}{\bd \Omega} - {\bd \Omega}{\bd A},
\nonumber
\eeq
where the skew symmetric tensor ${\bd \Omega}$ is called the spin. Xiao et al. 
\cite{Xiao98,Xiao98d} showed that an objective spin has the general form  
\beq{990}
{\bd \Omega}  = {\bd W} + \tens{P}({\bd V})\,  {\bd D} , 
\eeq
where ${\bd W}$ is the skew symmetric part of $\dot{\bd F}{\bd F}^{-1}$ and 
$ \tens{P}$ is an  isotropic fourth order tensor-valued function of ${\bd V}$. 
Lehmann and Liang \cite{Lehmann93} showed that using the rate associated with $\bd R$, i.e a corotational spin equal to the ``twirl'' ${\bd \Omega}^R=\dot{\bd R} {\bd R}^t$, the Eulerian and Lagrangian stresses conjugate to   $f({\bd V})$ and  $f({\bd U})$ are related  by ${\bd \tau} = {\bd R} {\bd T}{\bd R}^t$.   This relationship simply rotates the Lagrangian stress, but does not reproduce the Cauchy stress for any choice of $f$.   

The fundamental relation for Eulerian conjugate stress is based on the finding of
Xiao et al.  \cite{Xiao97} that 
\beq{-11}
\stackrel{\circ_\text{log} }{\ln ({\bd V} )}= {\bd D} ,
\eeq
where $\circ_\text{log} $ denotes an objective corotational rate defined by the logarithmic spin 
${\bd \Omega}^\text{log}$ \cite{Xiao97}.  We will discuss ${\bd \Omega}^\text{log}$ in detail, providing a new derivation and representation, and comparison with ${\bd \Omega}^R$.  The relationship \rf{-11} allows us to define a class of work-conjugate Eulerian stress-strain pairs for all $f({\bd V})$ that includes the Cauchy stress.   
However, it should be borne in mind that the logarithmic rate is but one from a continuum of possibilities. 

A second objective of this paper is a generalization of eq. \rf{-11} to arbitrary strain measure $f ({\bd V} )$.  Xiao et al.  \cite{{Xiao97}}  proved that ${\bd D}$ is recovered only from  the Hencky strain 
$\ln ({\bd V} )$ combined with the logarithmic spin; no other strain measure can yield   ${\bd D}$ no matter what spin is used.  Here we will show that for a given  strain measure $f ({\bd V} )$ there is a unique spin which provides the best \emph{approximation} to ${\bd D}$, and the corresponding conjugate stress is the best approximation to the Cauchy stress. 

\subsection{Summary of principal results}


  Our first main result  is: 
\begin{thm}\label{thm1}
The stress conjugate to the Eulerian strain $f  ({\bd V} )$ is
\beq{-12}
{\bd \tau}  = \big( \nabla f({\bd V}) \big)^{-1}\, 
{\bd \tau}^{(1)}({\bd \Omega}),
\eeq
where ${\bd \tau}^{(1)}$ depends on the corotational rate used, 
\beq{263}
{\bd \tau}^{(1)}({\bd \Omega})  =
\big[ {\bd V}^2\boxtimes{\bd I} + {\bd I} \boxtimes{\bd V}^2 + ({\bd V}^2\boxtimes{\bd I} - {\bd I} \boxtimes{\bd V}^2) \tens{P}({\bd V})\big]^{-1} ({\bd V}\boxtimes{\bd I} + {\bd I} \boxtimes{\bd V}) \, {\bd \sigma}
. \nonumber
\eeq
 \end{thm}
We will explain this result in detail, and provide alternative representations for eq. \rf{-12}.  We note at this stage the similarity in  form between  the relations \rf{-4} and \rf{-12}. In particular, the stress ${\bd \tau}^{(m)}$ conjugate to the strain $ f^{(m)}({\bd V})$ is 
\begin{subequations}\label{-01}
\bal{-01a}
{\bd \tau}^{(0)}&=   \big( \nabla \ln {\bd V} \big)^{-1} \, {\bd \tau}^{(1)} 
= \int_0^1 \dd x\, {\bd V}^{x} \boxtimes {\bd V}^{1 - x}\, {\bd \tau}^{(1)}, 
\\
{\bd \tau}^{(\frac13)}&= \frac{1}{3}  
  ({\bd V}^{2/3}\boxtimes{\bd I} +{\bd V}^{1/3}\boxtimes{\bd V}^{1/3}  +{\bd I} \boxtimes{\bd V}^{2/3}   )\,{\bd \tau}^{(1)} , \label{-01b1}
 \\
{\bd \tau}^{(\frac12)}&= \frac12 ( \sqrt{\bd V} \boxtimes{\bd I} + {\bd I}\boxtimes\sqrt{\bd V}
)\, {\bd \tau}^{(1)}, \label{-01b}
\\
{\bd \tau}^{(2)}&= 2 ( {\bd V} \boxtimes{\bd I} + {\bd I}\boxtimes{\bd V}
)^{-1}{\bd \tau}^{(1)},
\label{-01c}
  \\
{\bd \tau}^{(-m)}&= ({\bd V}^{m}\boxtimes{\bd V}^{m})\,{\bd \tau}^{(m)}.\label{-01d}
\end{align}
\end{subequations}

The second principal result defines a new spin defined by the Eulerian strain measure. 
\begin{thm}\label{thm3}
For every Eulerian strain measure $f({\bd V})$ there is a unique corotational rate which minimizes the difference between 
the conjugate stress and the Cauchy stress.  The rate is defined by the f-spin ${\bd \Omega}^f = {\bd W} + \tens{P}^f{\bd D}$ which depends upon the function $f$ via the fourth order projection tensor 
\beq{86-}
\tens{P}^f  =
 ( {\bd V}\boxtimes{\bd I} - {\bd I} \boxtimes{\bd V})^*
\big[ 
  \big( \nabla f({\bd V} ) \big)^{-1}
-( {\bd V}\boxtimes{\bd I} + {\bd I} \boxtimes{\bd V} )^{-1}( {\bd V}^2\boxtimes{\bd I} + {\bd I} \boxtimes{\bd V}^2 )
\big], \nonumber
\eeq
and $\tens{A}^*$ denotes the 
pseudo-inverse (or Moore-Penrose inverse) of the tensor $\tens{A}$.  The conjugate stress 
using the f-spin is 
\beq{220-}
{\bd \tau} = {\bd \sigma}^f \equiv {\bd \sigma}  + \sum\limits_{i=1}^n \big(\frac1{\lambda_i f'(\lambda_i)}  -1\big) {\bd V}_i \boxtimes{\bd V}_i\, {\bd \sigma},
\eeq
where  $\lambda_i$ are the principal stretches,   ${\bd V}_i$ the principal dyads, i.e. the eigenvalues and eigentensors of ${\bd V}$, and the eigen-index $n \in\{1,2,3\}$ is the number of distinct eigenvalues.   The conjugate stress is minimal in the sense that 
$|{\bd \tau} -{\bd \sigma}|>   |{\bd \sigma}^f- {\bd \sigma}|$ for any other corotational rate. 
\end{thm}
The pseudo-inverse is a unique quantity and will be defined in detail later.  

The logarithmic spin \cite{Xiao98c} is a very special case of the f-spin.  
It is clear from eq. \rf{220-} that when  $f(x)=\ln x$ \emph{and} the f-spin is used then the conjugate stress is simply the Cauchy stress, i.e.  ${\bd \sigma}^{\ln} = {\bd \sigma}$. 
Note that $\bd \sigma$  is recovered as ${\bd \tau}^{(0)}$, the stress conjugate to the Hencky strain $\ln {\bd V}$. 
  No other spin reproduces the Cauchy stress as the conjugate of any strain \cite{Xiao97}.   This emphasizes the mutual relation between the Hencky strain and the logarithmic spin.  

\subsection{Review and plan of the paper}

No attempt is made to summarize the considerable literature on work-conjugacy, strain measures and associated stresses, although two introductory reviews are worthy of  mention.   Curnier and Rakotomanana \cite{Curnier91} provide an instructive overview of strain measures and conjugate stresses, with extensive references to the literature prior to 1990.   A more concise but in-depth description of work conjugacy and its implications is given by Ogden \cite{Ogden84}.  These reviews and most of the  work   prior to 1991 dealt with stress conjugate to Lagrangian strain measures, although there had been some relevant work on quantities related to $\ln {\bd V}$.  For instance,  Fitzgerald \cite{Fitz80}  considered  the stress  conjugate to $\ln {\bd V}$ in the context of hyperelasticity.    Hoger \cite{Hoger87} derived expressions for the rate of change of ${\bd U}$, which subsequently proved useful for Lehmann et al. \cite{Lehmann91,Lehmann93} when they considered ${\bd V}$ specifically. 
  The focus here is on Eulerian strain and its work-conjugate stress, and builds upon developments in the 1990s. Lehmann and Liang  \cite{Lehmann93} introduced a clear procedure to extend the idea of work-conjugacy to strains whose rates are not objective in a fixed frame, see also \cite{Lehmann91}.  The idea, reviewed in Section \ref{sec4}, permits the use of corotational rates.  This is especially important for Eulerian strain measures since Xiao et al. \cite{Xiao97} proved that the only way to obtain ${\bd D}$ with Eulerian strain is  as the logarithmic rate of $\ln {\bd V}$.  
  We emphasize that  the pointwise rate of working $\dot{w}$ is the focus as we consider its implications for pointwise stress based on different definitions of strain.   No assumptions of material homogeneity, isotropy, or otherwise is assumed or required. 

Xiao, Bruhns and Meyers  provide  the most complete and thorough analysis of Eulerian conjugate stress and strain. In a series of groundbreaking papers \cite{Xiao97,Xiao98,Xiao98d} culminating in \cite{Xiao98c} these authors showed that the notion of conjugate stress is just as relevant to Eulerian strain as it is for Lagrangian strain.  Because  the role of the rate, or spin, is central to the Eulerian problem but is absent from Lagrangian work-conjugacy, it is essential to have a thorough understanding of the  possible spin tensors and their dependence on quantities such as  $\bd D$, $\bd W$ and  $\bd V$.   Once this is understood  then the form of the conjugate stress becomes apparent.  Xiao et al. \cite{Xiao98c} derived expressions for the Eulerian conjugate stress for arbitrary strain measures $f$ and for arbitrary permissible corotational strain rate.  Their subsequent  work has highlighted the role of the logarithmic rate and the Hencky strain 
in applications to hyperelasticity and other constitutive theories, see  \cite{Xiao06} for a thorough review. 

This paper presents  new results which extend the work of Xiao et al.  in several directions.  The introduction and discovery of the role of  the f-spin shows that there is a certain unique  conjugate  stress associated with every Eulerian  strain measure.  The dual formulation for the Eulerian and Lagrangian conjugate stresses in eqs. \rf{-4} and \rf{-12}   further emphasizes   the similarities in the two descriptions.  The formulation throughout is in direct tensor notation, which we believe makes the results more transparent. 

The plan of the paper is  as follows. The notation is introduced in Section \ref{sec2}, where  the  gradient and the pseudo-inverse of a tensor are defined.  Corotational strain rates are discussed in Section \ref{sec3} and some basic results for Eulerian strain measures are derived.   The f-spin is  introduced and discussed in Section \ref{sec4}. It is shown that the corotational rate defined by the f-spin, or f-rate, has certain unique and desirable properties.   The main results for conjugate stress-strain pairs are deduced in Section \ref{sec5}.

\section{Tensors functions  and the pseudo-inverse.}\label{sec2}

\subsection{Preliminaries}

We will be dealing with tensors of second and fourth order. 
Second order tensors act  on vectors in 
a three dimensional inner product space,  
 ${\bd x} \rightarrow {\bd A}{\bd x}$ with  transpose ${\bd A}^t$ such that 
 ${\bd y}\cdot{\bd A}{\bd x} = {\bd x}\cdot{\bd A}^t{\bd y}$.  Spaces of symmetric and skew-symmetric tensors are distinguished, Lin = Sym $\oplus$ Skw 
where   ${\bd A}\in $ Sym (Skw) iff 
${\bd A}^t={\bd A}$ (${\bd A}^t=-{\bd A}$).  The inner product on Lin is defined by 
${\bd A}\cdot {\bd B} = \tr({\bd A} {\bd B}^t)$.   Products ${\bd A} {\bd B}\in$Lin are defined by 
${\bd y}\cdot{\bd A} {\bd B}{\bd x} =   ({\bd A}^t{\bd y}) \cdot {\bd B}{\bd x}$. 

Psym  is the space of positive definite second order tensors.
When dealing  with functions of a symmetric tensor it is often useful to 
rephrase the functional form  in terms of the  spectral decomposition: 
\beq{-71}
{\bd A}  = \sum\limits_{i=1}^n \alpha_i {\bd A}_i ,  
\qquad 
{\bd I}  = \sum\limits_{i=1}^n  {\bd A}_i ,  
\qquad  
{\bd A}_i{\bd A}_j = \begin{cases}
{\bd A}_i & i=j,\\
0 , & i\ne j,  
\end{cases}
\eeq
where ${\bd A}_i\in$Psym and $n \le 3$ is the eigen-index. 
Thus, 
\beq{883}
f( {\bd A}) =   \sum\limits_{i=1}^n f(\alpha_i) {\bd A}_i . 
\nonumber
\eeq
The Poisson bracket of two second order tensors is 
\beq{-8}
\big\{ {\bd A},  {\bd B} \big\} = {\bd A}   {\bd B}-{\bd B}{\bd A}. 
\nonumber
\eeq

${\tens L}$in is the space of fourth order tensors acting on Lin,  
 ${\bd X} \rightarrow {\tens A}{\bd X}$ with  transpose ${\tens A}^t$ such that 
 ${\bd Y}\cdot {\tens A}{\bd X} = {\bd X}\cdot{\tens A}^t{\bd Y}$ for all
 ${\bd X}$, ${\bd Y}\in$Lin.  The vector space may be decomposed ${\tens L}$in $= {\tens S}$ym$ \oplus  {\tens S}$kw 
where  ${\tens S}$ym and ${\tens S}$kw denote the spaces of symmetric (${\tens A}^t={\tens A}$) and skew-symmetric (${\tens A}^t=-{\tens A}$) tensors, respectively.   
Any ${\tens A}\in  {\tens L}$in can be uniquely partitioned into symmetric and skew parts: 
${\tens A}= {\tens A}^{(+)}+{\tens A}^{(-)}$, where 
${\tens A}^{(\pm)} = ({\tens A}\pm {\tens A}^t)/2$. 
 The identity ${\tens I}$ satisfies ${\tens I}{\bd X} = {\bd X} $ for all ${\bd X}\in$ Lin.   The product ${\tens A} {\tens B}\in \tens{L}$in is defined by 
${\bd Y}\cdot{\tens A} {\tens B}{\bd X} =   ({\tens A}^t{\bd Y}) \cdot {\tens B}{\bd X}$.   
${\tens P}$sym  is the space of positive definite fourth order tensors:  ${\tens A}\in {\tens P}$sym  iff ${\bd X}\cdot {\tens A}{\bd X} >0$, for all nonzero $ {\bd X} \in$ Sym.  

The square tensor product ${\bd X}\boxtimes {\bd Y}$, Lin$\times$Lin$\rightarrow {\tens L}$in, is defined  by  \cite{Rosati00}
\beq{51}
( {\bd X}\boxtimes {\bd Y}) {\bd Z} = {\bd X}{\bd Z}{\bd Y}^t, 
\qquad \forall \,  {\bd Z}\in \text{ Lin}. 
\nonumber
\eeq
In particular, we note the property
\beq{53}
( {\bd A}\boxtimes {\bd B}) ( {\bd X}\boxtimes {\bd Y})    = 
( {\bd A}{\bd X}) \boxtimes ({\bd B}{\bd Y}) .
\nonumber
\eeq

\subsubsection{The tensor gradient function  and its inverse}

The gradient of a tensor function $f({\bd A})$  is a fourth order tensor $\nabla f \in \tens{L}$in defined by 
\beq{44}
 \nabla f ({\bd A}) \, {\bd X} = 
 \lim\limits_{\epsilon \rightarrow 0} \frac1{\epsilon }\big[ 
 f ({\bd A} +\epsilon {\bd X}) - f ({\bd A})
 \big]. 
 \eeq
We make extensive use of the following representation 
which uses  the spectral form of ${\bd A}$,  
\beq{45}
 \nabla f ({\bd A})   = 
 \sum\limits_{i,j=1}^n \, \frac{ f(\alpha_i) - f(\alpha_j)}{ \alpha_i - \alpha_j} 
 {\bd A}_i \boxtimes {\bd A}_j, 
 \nonumber
 \eeq
 where the ratio becomes $f'(\alpha_i)$ for $i=j$.  
 The formula \rf{44} for the first derivative is well known, e.g.  \cite{Ogden84,Xiao95}. Norris \cite{Norris07b}   provides formulas for the $n^{th}$ derivative of a tensor valued-function.  We define the inverse tensor function 
  $ \triangle f ({\bd A}) \in \tens{L}$in by 
\beq{456}
 \triangle f ({\bd A})  \equiv \big(\nabla f ({\bd A})\big)^{-1} = 
 \sum\limits_{i,j=1}^n \, \frac{ \alpha_i - \alpha_j}{ f(\alpha_i) - f(\alpha_j)} 
 {\bd A}_i \boxtimes {\bd A}_j , 
 \nonumber
 \eeq
where the ratio is $1/f'(\alpha_i)$ for $i=j$. The definition of $ \triangle f ({\bd A})$ is problematic if $f'(\alpha_i)$ vanishes, but we preclude this possibility next by restricting consideration to strictly monotonic functions: strain measure functions.

\subsection{Strain measure functions}
 
 The function $f$ is a  \emph{strain measure} \cite{Hill78,Scheidler91} if it is a smooth function 
$ f : \tens{R}^+ \rightarrow \tens{R}$ which satisfies 
\beq{004}
f(1)=0,\qquad f'(1)=1, \qquad f'>0. 
\nonumber
\eeq
It may be shown  \cite{Norris07b} that 
the gradient of a strain measure function and its inverse are positive definite fourth order tensors, i.e. 
$\nabla f ({\bd A}),\, \triangle f ({\bd A}) \in \tens{P}$sym.
We restrict attention to strain measure functions 
for the remainder of the paper.

 \subsection{The pseudo-inverse}
  
  For ${\bd A}\in \text{Psym}$ consider the equation 
  \beq{-55}
  \big\{ {\bd A},  {\bd X} \big\} = {\bd Y},
  \eeq
for the unknown ${\bd X}$ in terms of  ${\bd Y}$.  It is assumed that ${\bd Y}$    is either symmetric or skew  and ${\bd X}$ is of the opposite parity      \cite{Dui06}. 
The equation can be written ${\bd A}{\bd X} - {\bd X} {\bd A} = {\bd Y}$, or 
\beq{61}
\tens{J}({\bd A}){\bd X}  = {\bd Y} 
\quad
\text{where  } \quad 
\tens{J}({\bd A})  \equiv {\bd A}\boxtimes{\bd I} - {\bd I} \boxtimes{\bd A}.
\nonumber
\eeq
We will only consider $\tens{J}({\bd A})$ for symmetric $\bd A$, implying 
$\tens{J}\in \tens{S}$ym and $\tens{J}$ maps Sym$\rightarrow$Skw and Skw$\rightarrow$Sym. Therefore, 
$\tens{J}$ does not possess eigenvalues,  eigenvectors or an inverse in the usual sense.  

The unique solution of the tensorial equation \rf{-55} is \cite{Norris07b}
 \beq{-56}
 {\bd X} = \tens{J}^*({\bd A}){\bd Y} = ({\bd A}\boxtimes{\bd I} - {\bd I} \boxtimes{\bd A})^* {\bd Y}. 
  \eeq 
The pseudo-inverse, or equivalently the Moore-Penrose inverse,  $\tens{J}^*$,  is  defined such that 
\beq{631}
\tens{J}\tens{J}^*\tens{J}   = \tens{J},
 \qquad
\tens{J}^*\tens{J} \tens{J}^*  = \tens{J}^* . 
\eeq
The spectral forms of $\tens{J}({\bd A})$ and its pseudo-inverse are
\beq{66}
\tens{J}({\bd A})  = \sum\limits_{i,j=1}^n  (\alpha_i - \alpha_j) {\bd A}_i\boxtimes{\bd A}_j ,
\qquad
\tens{J}^*({\bd A})  =  \sum\limits_{\substack{i,j=1 \\ i\ne j}}^n \, 
 (\alpha_i - \alpha_j)^{-1} {\bd A}_i\boxtimes{\bd A}_j . 
 \nonumber
\eeq
These  clearly satisfy \rf{631}.  

Further insight into the pseudo-inverse is gained 
by introducing the set of $N \le 6$ fourth order tensors   
associated with ${\bd A}\in$Sym, 
\beq{038}
 \tens{A}_I =  \begin{cases} 
    \bd{A}_I\boxtimes \bd{A}_I, &  I=1,\ldots , n, 
  \\
   \bd{A}_i\boxtimes \bd{A}_j+\bd{A}_j\boxtimes \bd{A}_i ,
 & I=n+1, \ldots N. 
  \end{cases} 
  \eeq
$N=6$ for    $n=3$  and the indices  $I=4,5,6$ correspond to $(i,j) = (2,3),(3,1),(1,2)$, respectively.   Similarly $N=3$ if  $n=2$ and $N=1$ if $n=1$. 
Note that
\beq{-99}
{\tens I} = \sum\limits_{I=1}^N {\tens A}_I, 
\qquad  
{\tens A}_I{\tens A}_J = \begin{cases}
{\tens A}_I & I=J,\\
0 , & I\ne J .  
\end{cases}
\eeq
The identity ${\tens I}   =  {\bd I} \boxtimes {\bd I}$  implies 
the partition of unity $\rf{-99}_1$, and it may be readily checked that the $\tens{A}_I $ satisfy the orthogonality conditions $\rf{-99}_2$.  

The pseudo-inverse satisfies 
\beq{67}
\tens{J}^*\tens{J} = \tens{J}\tens{J}^* = 
\tens{I} - \sum\limits_{I =1}^n   {\tens A}_I  = \sum\limits_{I =n+1}^N   {\tens A}_I .  
\eeq
This is never equal to the identity $\tens{I} $, 
which is the property that distinguishes the pseudo-inverse from the standard notion of inverse. 
Further properties of the pseudo-inverse are presented in Norris \cite{Norris07b}.   The explicit solution of eq. \rf{61} can be expressed in a variety of ways without the use of fourth order tensors. Perhaps the simplest is  the recently discovered solution of Dui et al. \cite{Dui07}, 
\beq{-456}
{\bd X} = \big( 3 {{\bd A}'}^2 - \frac12 (\tr {{\bd A}'}^2){\bd I}\big)^{-1}
 ( 2 {\bd A}'{\bd Y} + {\bd Y}{\bd A}'), 
 \nonumber
 \eeq
  where ${\bd A}'$ is the deviatoric part of ${\bd A}$.

\section{Kinematics}\label{sec3}

\subsection{Basics}

The polar decomposition of the deformation gradient is ${\bd F} = {\bd R}{\bd U}= {\bd V}{\bd R}$  where ${\bd R}\in$SO(3) satisfies ${\bd R}{\bd R}^t = {\bd R}^t{\bd R}= {\bd I}$ 
and the right and left stretch tensors ${\bd U}$ and ${\bd V}$ are positive definite and related by 
${\bd V} = ({\bd R}\boxtimes {\bd R}){\bd U}$.    The fundamental Eulerian strain can be taken as either $\bd V$ or 
its square $
{\bd B}  = 
{\bd V}^2 = {\bd F}{\bd F}^t$.   
The spectral representations of ${\bd V}$ and ${\bd B}$ are
\beq{72}
{\bd V}  = \sum\limits_{i=1}^n \lambda_i {\bd V}_i ,
\qquad
{\bd B}  = \sum\limits_{i=1}^n \beta_i {\bd V}_i , 
\qquad 
{\bd V}_i{\bd V}_j = \begin{cases}
{\bd V}_i, & i=j,\\
0 , &i\ne j,  
\end{cases}
\eeq
where $\lambda_i>0$ and $\beta_i = \lambda_i^2$. 

The rate of change of ${\bd B}$ is  $\dot{\bd B}= {\bd L} {\bd B} + {\bd B}{\bd L}^t$ where
${\bd L} = \dot{{\bd F}}{\bd F}^{-1}$. %
Let  ${\bd D}\in$Sym and  ${\bd W}\in$Skw be the symmetric and skew-symmetric parts of ${\bd L}$, respectively.  Thus, 
${\bd L}= {\bd D} + {\bd W}$ and $\dot{\bd B}$ can be expressed 
\beq{4}
\dot{\bd B} =  ({\bd I}\boxtimes{\bd B}-{\bd B}\boxtimes{\bd I}){\bd W}
+ ({\bd I}\boxtimes{\bd B}+{\bd B}\boxtimes{\bd I}){\bd D}. 
\eeq
We will find this form useful for deriving  more general  strain rates. 

\subsection{Co-rotational rates}

  Let ${\bd A}(t)$ be a symmetric second order tensor,
and ${\bd \Omega}$ is skew and arbitrary.   
Define  the  corotational rate 
\beq{-1}
\stackrel{\circ}{{\bd A}} \equiv \dot{\bd A} + \{ {\bd A}, {\bd \Omega}\},
\qquad {\bd \Omega} \in \text{Skw}. 
\eeq
  For any $ {\bd \Omega}(t)\in$Skw we can identify a
rotation ${\bd Q}(t)\in $SO(3) such that 
\beq{-14}
\dot{ \overline{ {\bd Q} {\bd A}{\bd Q}^t }} = 
{\bd Q} \stackrel{\circ}{{\bd A}} {\bd Q}^t
. \nonumber
\eeq
Differentiating the left member and using \rf{-1} for the right member implies that 
${\bd \Omega} = -{\bd Q}^t \dot{\bd Q} $.   Hence, 
${\bd Q}$ must satisfy $\dot{\bd Q} = -{\bd Q}{\bd \Omega}$, with solution 
unique up to a rigid body rotation.  
The corotational rate may therefore be interpreted as the Lie derivative with respect to spatial rotation defined by 
${\bd Q}(t)$.   Thus, let $\phi$ define the mapping (rotation) ${\bd x} \rightarrow {\bd Q}{\bd x}$, then 
the corotational rate is $\phi [ \frac{\dd}{\dd t} \phi^{-1} (\cdot )]$. 

The Jaumann rate $\stackrel{\circ_J }{{\bd A}}$   defined by $ {\bd \Omega} = {\bd W}$ corresponds to $ \tens{P} = 0$ in 
eq. \rf{990}.   Using the latter formula to parameterize the  spin $ {\bd \Omega}$ allows us to express the general corotational of ${\bd A}$ as 
\beq{295}
\stackrel{\circ }{{\bd A}} = 
\stackrel{\circ_J }{{\bd A}} + 
({\bd A}\boxtimes{\bd I} - {\bd I} \boxtimes{\bd A}) \tens{P}({\bd V})\,  {\bd D} . 
\eeq
Equation \rf{4} implies that the Jaumann rate of ${\bd B}$ is $\stackrel{\circ_J }{{\bd B}} 
=({\bd I}\boxtimes{\bd B}+{\bd B}\boxtimes{\bd I}){\bd D}$.   The general rate $\stackrel{\circ }{{\bd B}}$ then follows from 
\rf{295}, and $\stackrel{\circ }{{\bd V}}$ can  be determined from the identity 
$\stackrel{\circ }{{\bd B}} = ({\bd V}\boxtimes{\bd I} + {\bd I} \boxtimes{\bd V})\stackrel{\circ }{{\bd V}}$.  
In summary, the general form of the corotational rate of the fundamental Eulerian strains are 
\begin{subequations}
\bal{296}
\stackrel{\circ }{{\bd B}} &= 
\big[ {\bd B}\boxtimes{\bd I} + {\bd I} \boxtimes{\bd B} +
({\bd B}\boxtimes{\bd I} - {\bd I} \boxtimes{\bd B}) \tens{P}({\bd V})\big]\,  {\bd D} ,
  \\
\stackrel{\circ }{{\bd V}} &= 
\big[({\bd V}\boxtimes{\bd I} + {\bd I} \boxtimes{\bd V})^{-1}  ({\bd V}^2\boxtimes{\bd I} + {\bd I} \boxtimes{\bd V}^2)+
({\bd V}\boxtimes{\bd I} - {\bd I} \boxtimes{\bd V}) \tens{P}({\bd V})\big]\,  {\bd D} .
\label{296b}
\end{align}
\end{subequations}

\subsection{Spins} 

Many   candidates have been considered from the infinity of possible spins  \cite{Dill06}.  For instance, the polar spin 
\beq{404}
{\bd \Omega}^R=\dot{\bd R} {\bd R}^t,
\eeq
corresponding to ${\bd Q} = {\bd R}^t$, is  useful  as a comparison spin.   
Other common spins \cite{Xiao98} are ${\bd \Omega}^E$  defined by the ``twirl" of  the Eulerian principal axes
and ${\bd \Omega}^L$ related to the  Lagrangian principal axes.  
It is shown in  Appendix \ref{appb} that 
${\bd \Omega}^\alpha ={\bd W} + \tens{P}^\alpha {\bd D}$, $\alpha = R, E, L$, 
where 
\begin{subequations}\label{-406}
\bal{-406a}
\tens{P}^R &
= (  {\bd I} \boxtimes{\bd V} -{\bd V}\boxtimes{\bd I} )
(  {\bd I} \boxtimes{\bd V} +{\bd V}\boxtimes{\bd I} )^{-1}, 
\\ 
\tens{P}^E &  
= (  {\bd I} \boxtimes{\bd V}^2 -{\bd V}^2\boxtimes{\bd I} )^*
(  {\bd I} \boxtimes{\bd V}^2 +{\bd V}^2\boxtimes{\bd I} ),
\label{-406b}
\\ 
\tens{P}^L &
= (  {\bd I} \boxtimes{\bd V}^2 -{\bd V}^2\boxtimes{\bd I} )^*
\, 2   {\bd V} \boxtimes{\bd V}.
\label{-406c}
\end{align}
\end{subequations}
The three spins ${\bd \Omega}^R$, ${\bd \Omega}^E$ and ${\bd \Omega}^L$ are related by 
${\bd \Omega}^E - {\bd \Omega}^R = {\bd \Omega}^L - {\bd W}$, see 
Appendix \ref{appb}.  The fourth order projection tensors are therefore connected by 
$ \tens{P}^E - \tens{P}^L= {\tens{P}^R}$,  and we note the additional relation $ \tens{P}^E + \tens{P}^L= {\tens{P}^R}^*$, which is readily verified.

The  most general form of the  isotropic tensor-valued function $ \tens{P}\in {\tens S}$ym   
 involves three isotropic scalar functions $\nu_1,\nu_2, \nu_3$   \cite{Xiao98},
\bal{291}
  \tens{P}({\bd V}) &= ({\bd V}\boxtimes{\bd I} - {\bd I}\boxtimes{\bd V} )
  \big[ \nu_1 {\tens I}  + \nu_2 ({\bd V}\boxtimes{\bd I} + {\bd I}\boxtimes{\bd V} )
  + \nu_3 {\bd V}\boxtimes{\bd V}
  \big]  
  \nonumber \\
   & = 
  \sum\limits_{\substack{i,j=1 \\ i\ne j}}^n p_{ij}  \, {\bd V}_i\boxtimes{\bd V}_j ,
\end{align}
where
\beq{292}
  p_{ij}  =(\lambda_i - \lambda_j) \big[ \nu_1 + (\lambda_i + \lambda_j)\nu_2 +\lambda_i\lambda_j \nu_3\big] ,
 \qquad \nu_k=\nu_k(I_1,I_2,I_3), \quad k=1,2,3.  
 \nonumber 
\eeq
Here, $I_1,I_2,I_3$ are the invariants of ${\bd V}$:  $I_1 = \tr {\bd V}$, $I_2 = \frac12 I_1^2 - \frac12\tr {\bd V}^2$, $I_3 = \det {\bd V}$. 

The corotational rate of $\bd V$ can now be written   
\beq{2961}
\stackrel{\circ }{{\bd V}} = \tens{Q} {\bd D}, 
\eeq 
where the fourth order tensor 
$\tens{Q} \in \tens{S}$ym follows from \rf{296b},  
\bal{-336}
\tens{Q} &= ({\bd V}\boxtimes{\bd I} + {\bd I} \boxtimes{\bd V})^{-1}  ({\bd V}^2\boxtimes{\bd I} + {\bd I} \boxtimes{\bd V}^2)+
({\bd V}\boxtimes{\bd I} - {\bd I} \boxtimes{\bd V}) \tens{P}({\bd V}) 
\nonumber  \\
&= 
\sum\limits_{i,j=1 }^n q_{ij}  \, {\bd V}_i\boxtimes{\bd V}_j  ,
\nonumber 
\end{align}
and 
\beq{-+-}
q_{ij} = (\lambda_i - \lambda_j) p_{ij} + \frac{\lambda_i^2+ \lambda_j^2}{\lambda_i + \lambda_j}. 
\eeq

\subsection{Eulerian strain measures}

The Lagrangian Seth-Hill  strain ${\bd E}^{(m)} = m^{-1} ( {\bd U}^m - {\bd I})$ corresponds to 
$f(x) = f^{(m)}(x)$.  We  define the analogous Eulerian strain
\beq{73}
{\bd e}^{(m)} = f^{(m)} ({\bd V}) =  m^{-1} ( {\bd V}^m - {\bd I}), 
\eeq
and note in particular the Hencky strain ${\bd e}^{(0)} = \ln {\bd V}$. 
Other examples include
\beq{75}
{\bd e}^{(1)} = {\bd V}-{\bd I},  
\quad
{\bd e}^{(2)} = \frac12({\bd B}-{\bd I}),  
\quad
{\bd e}^{(-1)} = {\bd I}-{\bd V}^{-1},  
\quad
{\bd e}^{(-2)} = \frac12( {\bd I}-{\bd B}^{-1}). 
\nonumber 
\eeq

\subsection{Eulerian strain rates}

We now present some identities  for the corotational  rates of Eulerian strains.  
These will prove useful later in deriving conjugate Eulerian stresses.  The 
first identity  applies to arbitrary strain measures:  
\begin{lem}\label{lem-1}
The corotational rate of any Eulerian strain measure $f({\bd V})$ is 
\beq{+7}
\stackrel{\circ }{f({\bd V})}
= [\nabla f({\bd V})]\, \tens{Q}({\bd V}) \, {\bd D}. 
\nonumber 
\eeq
\end{lem}
The proof is a simple application of the chain rule using eq. \rf{2961} for $\stackrel{\circ }{{\bd V}}$.  This separates the dependence on the strain measure $f$ from the dependence on the particular corotational rate used, which determines $ \tens{Q}$. 

The second identity connects the strain rate with the Hencky strain: 
\begin{lem}\label{lem4}
The strain rate of any Eulerian strain measure $f({\bd V})$ can be expressed in terms of the Hencky strain rate as
\beq{891}
\stackrel{\circ }{f({\bd V})}
= [\nabla f({\bd V})]\, (\triangle \ln {\bd V}) \, \stackrel{\circ }{\ln{\bd V}}. 
\nonumber 
\eeq
\end{lem}
The proof is a straightforward generalization of the chain rule of differentiation \cite[Thm. 2]{Xiao98}. Let  ${\bd M} =\ln{\bd V}$ and $f({\bd V}) = \hat{f}({\bd M})$ then, 
\beq{-91}
\stackrel{\circ }{\hat{f}({\bd M})}
= \nabla_{\bd M} \hat{f}({\bd M})\, \stackrel{\circ }{\bd M}
=  \nabla f({\bd V})\, ( \nabla_{\bd M} {\bd V} )\, \stackrel{\circ }{\bd M}.  
\nonumber 
\eeq
But the fourth order tensor $\nabla_{\bd M} {\bd V}$ is just the inverse of $\nabla_{\bd V} {\bd M}$
since $
\nabla_{\bd V} {\bd V} = \tens{I}$.

\section{The f-spin and the  logarithmic spin }\label{sec4}

\subsection{Strain rate and the stretching tensor}
In order to 
make the connection between the kinematics and the power $\dot{w}$ we must relate some strain rate to the stretching tensor $\bd D$.   A general connection can be found by starting with the rate of change of  an  arbitrary tensor valued function of ${\bd B}$.  Thus,    
\beq{817}
\dot{\tilde{f}}( {\bd B}) = 
[\nabla \tilde{f} ({\bd B })] \dot{\bd B} = 
\sum\limits_{i,j=1}^n \frac{ \tilde{f}(\beta_i)- \tilde{f}(\beta_j)}{\beta_i-\beta_j}\, {\bd V}_i \boxtimes{\bd V}_j \dot{\bd B} , \nonumber 
\eeq
where  the temporary definition $\tilde{f}(x) = f(x^2)$ is used, so that $f ({\bd V})=\tilde{f}({\bd B})$. 
Reverting to $f ({\bd V})$ and  using $\beta_i=\lambda_i^2$, the rate of change of the associated function of ${\bd V}$ is 
\beq{81}
\dot{f}( {\bd V}) =  
\sum\limits_{i,j=1}^n \frac{ f(\lambda_i)- f(\lambda_j)}{\lambda_i^2-\lambda_j^2}\, {\bd V}_i \boxtimes{\bd V}_j \dot{\bd B} ,
\eeq
where the ratio becomes $f'(\lambda_i)/(2\lambda_i) $ for $i=j$.   Substituting $\dot{\bd B}$  into \rf{81} 
 and using the filtering properties of ${\bd V}_i$, such as 
$({\bd V}_i \boxtimes{\bd V}_j)({\bd I}\boxtimes{\bd B}) = \beta_j {\bd V}_i \boxtimes{\bd V}_j$, gives
\bal{844}
 \dot{f }({\bd V}) 
&=  \sum\limits_{i,j=1}^n  \big( f(\lambda_j)- f(\lambda_i)\big) {\bd V}_i  \boxtimes{\bd V}_j {\bd W}
+ \sum\limits_{i,j=1}^n  \frac{ f(\lambda_i)- f(\lambda_j)}{\beta_i-\beta_j}\,(\beta_i+\beta_j)
{\bd V}_i \boxtimes {\bd V}_j  \, {\bd D}
\nonumber \\
&= \big\{ {\bd W}, f({\bd V}) \big\}
+ \sum\limits_{i,j=1}^n  \frac{ \lambda_i^2+\lambda_j^2}{\lambda_i^2-\lambda_j^2}  \big( f(\lambda_i)- f(\lambda_j)\big)\, 
{\bd V}_i \boxtimes {\bd V}_j  \, {\bd D}. 
\nonumber 
\end{align}
Adding and subtracting terms, this becomes 
\beq{845}
 \dot{f }({\bd V}) 
=  \widehat{\bf D} + \big\{ {\bd W}, f({\bd V}) \big\}
+ \sum\limits_{\substack{i,j=1 \\ i\ne j}}^n  \big[
\frac{ \lambda_i^2+\lambda_j^2}{\lambda_i^2-\lambda_j^2}   - \frac{1}{f(\lambda_i)- f(\lambda_j)}
\big]
 \big( f(\lambda_i)- f(\lambda_j)\big)\, 
{\bd V}_i \boxtimes {\bd V}_j  \, {\bd D}, 
\eeq
where $\widehat{\bd D}$ is a modified version of the stretching tensor, 
\beq{846}
 \widehat{\bd D} = {\bd D} 
+ \sum\limits_{i=1}^n  \big[
\lambda_i f'(\lambda_i)- 1
\big] 
{\bd V}_i \boxtimes {\bd V}_i  \, {\bd D}.  
\eeq
Note that the double sum in \rf{845} excludes the $i=j$ terms. 
We can therefore rewrite it in a form suggestive of a new corotational rate, 
\beq{847}
 \dot{f} ({\bd V})
 = \widehat {\bd D} + \big\{ {\bd \Omega}^f, f({\bd V}) \big\},
 \qquad  \qquad {\bd \Omega}^f  = {\bd W} + \tens{P}^f  {\bd D}, 
\eeq  
where   ${\bd \Omega}^f \in$Skw is called the f-spin, and its fourth order projection tensor is 
\beq{825}
 \tens{P}^f  = 
  ( {\bd V}\boxtimes{\bd I} - {\bd I} \boxtimes{\bd V})^*
\big[ 
  \triangle f( {\bd V} ) 
-  ( {\bd V}\boxtimes{\bd I} + {\bd I} \boxtimes{\bd V})^{-1}( {\bd V}^2\boxtimes{\bd I} + {\bd I} \boxtimes{\bd V}^2)  
\big]. 
\eeq
Alternatively, $ \tens{P}^f $ can be expressed in the form \rf{291} with matrix elements
\beq{0673}
p_{ij}^f =
\frac{1}{f(\lambda_i)- f(\lambda_j)} - 
\frac{ \lambda_i^2+\lambda_j^2}{\lambda_i^2-\lambda_j^2}  .
\nonumber 
\eeq
Note that $ {\bd \Omega}^f$ can blow up, but the action $\big\{ {\bd \Omega}^f, g({\bd V}) \big\}$ remains finite for any differentiable function $g$, including $f$.  In particular, the f-spin is an objective  material spin in the sense defined by Xiao et al. \cite{Xiao98}. 

The corotational rate associated with the f-spin is  defined in the usual manner as 
$\stackrel{\circ_f }{g({\bd V})} = 
 \dot{g}({\bd V}) + \big\{  g({\bd V}),{\bd \Omega}^f \big\}$.   The reason for introducing  this new rate is  
\beq{-44}
\stackrel{\circ_f }{f({\bd V})} 
=  \widehat {\bd D} , 
\nonumber 
\eeq
which  follows from eq. \rf{847}. 
This shows that for a particular choice of spin the corotational rate of an arbitrary strain measure $f({\bd V})$ is  related to the modified stretching tensor $ \widehat {\bd D}$.  The important point is that this is the closest, in a sense to be defined,  the strain rate can get to the actual  stretching tensor ${\bd D}$.    These ideas are made concrete through:
\begin{lem}\label{lem14}
 For any  objective corotational rate 
 \beq{-092}
|\stackrel{\circ }{f({\bd V})}  -   {\bd D}|^2 =  |\stackrel{\circ }{f({\bd V})} -   \widehat{\bd D}|^2 + |\widehat {\bd D} -   {\bd D}|^2 , 
\eeq
where $\widehat {\bd D}$ is the modified stretching tensor defined by eq. \rf{846}. 
\end{lem}
The proof follows by writing  
\bal{-45}
\stackrel{\circ }{f({\bd V})}  - {\bd D} 
&=
\stackrel{\circ }{f({\bd V})}  - \widehat {\bd D} + (\widehat {\bd D} -{\bd D} )
=\stackrel{\circ }{f({\bd V})} - \stackrel{\circ_f }{f({\bd V})}    + (\widehat {\bd D} -{\bd D} )
 \nonumber \\
 &=
 \sum\limits_{\substack{i,j=1 \\ i\ne j}}^n  (p_{ij} - p_{ij}^f ) \big( f(\lambda_i)- f(\lambda_j)\big) 
{\bd V}_i \boxtimes {\bd V}_j  \, {\bd D}
 + \sum\limits_{i=1}^n  \big[
\lambda_i f'(\lambda_i)- 1
\big] 
{\bd V}_i \boxtimes {\bd V}_i  \, {\bd D}. 
\end{align}
 Hence, 
 \beq{-0923}
|\stackrel{\circ }{f({\bd V})}  -   {\bd D}|^2 =  
\sum\limits_{\substack{i,j=1 \\ i\ne j}}^n  \big[(p_{ij} - p_{ij}^f )  \big( f(\lambda_i)- f(\lambda_j)\big)  \tr ( {\bd V}_i{\bd D}{\bd V}_j )\big]^2 
+  \sum\limits_{i=1}^n \big[ 
\big( \lambda_i f'(\lambda_i)- 1\big) \tr ( {\bd V}_i{\bd D})\big]^2 ,
\nonumber
\eeq
 where the two sums  on the right hand side are the corresponding terms in \rf{-092}.

 Therefore
\begin{lem}\label{lem1}
For every Eulerian strain measure $f$ there is a unique spin which minimizes the difference between 
$\stackrel{\circ }{f({\bd V})}$ and $\bd D$, and that spin is ${\bd \Omega}^f$.  The minimal difference is 
\beq{08}
|\stackrel{\circ_f }{f({\bd V})}-   {\bd D}|^2 = \sum\limits_{i=1}^n \big[ 
\big( \lambda_i f'(\lambda_i)- 1\big) \tr ( {\bd V}_i{\bd D})\big]^2.
\eeq  
\end{lem}
The proof follows using Lemma \ref{lem14} in the form  
\beq{093}
|\stackrel{\circ }{f({\bd V})} -   {\bd D}|^2 \ge |\widehat {\bd D} -   {\bd D}|^2 ,
\eeq
with equality iff ${\bd \Omega} = {\bd \Omega}^f$.

\subsection{The logarithmic spin}

Lemma \ref{lem1} implies that the corotational rate of strain equals $\bd D$  if 
 the strain measure has the property $x f'(x)-1 = 0$. The only solution satisfying the condition $f(1) =0$ is $f(x) = \ln x$, and the associated spin follows from \rf{825} as $ {\bd \Omega}^\text{log} = {\bd W} +  \tens{P}^\text{log} {\bd D} $ where 
\beq{87}
p_{ij}^\text{log}  =  
 \frac{1}{\ln\lambda_i- \ln\lambda_j} - 
\frac{\lambda_i^2+ \lambda_j^2}{\lambda_i^2- \lambda_j^2} .
\eeq
 $ {\bd \Omega}^\text{log}$ is the well known logarithmic spin  \cite{Xiao97}. 
Hence, of all possible rates and of all possible Eulerian strain measures  only the combination of the Hencky strain 
 and the rate defined by the logarithmic spin together  yield the strain rate ${\bd D}$.  This is  the unique relationship between $\ln {\bd V}$, ${\bd D}$ and ${\bd \Omega}^\text{log}$ which makes both the Hencky strain and the logarithmic spin special.   This result was first derived by  Xiao et al. \cite{Xiao97} and may be summarized as: 
\begin{lem}\label{lem6}
The strain rate $\bd D$ is recovered only as the corotational rate of the Eulerian strain  ${\bd e}^{(0)} = \ln {\bd V}$ with spin ${\bd \Omega}^\text{log}$ 
where 
the fourth order projection tensor   
$\tens{P}^\text{log} $  is given by eq. \rf{825} with $f=\ln$. 
That is, 
\beq{804} 
\stackrel{\circ_\text{log} }{ \ln {\bd V}}  = {\bd D} .
\eeq
\end{lem}

 \subsubsection{Some properties of the logarithmic spin}
 
An instructive alternative form for $\tens{P}^\text{log} $ is obtained by introducing 
 \beq{-433}
 \tens{P}^{\ln} \equiv 
 ( {\bd V}\boxtimes{\bd I} - {\bd I} \boxtimes{\bd V})^*
  \triangle \ln( {\bd V} )  ,  
 \eeq
so that  
\beq{-46}
\tens{P}^\text{log} = \tens{P}^{\ln}  + \tens{P}^E = \tens{P}^{\ln} +  \tens{P}^L +  \tens{P}^R.
\eeq
 Each of the projection tensors  may be expressed in terms of matrix elements $p_{ij}= - p_{ji}$ according to eq. \rf{291}: 
 \beq{-57}
p_{ij}^R = -   \frac{\lambda_i - \lambda_j}{\lambda_i + \lambda_j},
\qquad
p_{ij}^E  =  - \frac{\lambda_i^2+ \lambda_j^2}{\lambda_i^2- \lambda_j^2} ,
\qquad 
p_{ij}^L  =  - \frac{2\lambda_i  \lambda_j}{\lambda_i^2- \lambda_j^2} ,
\qquad 
p_{ij}^{\ln}  =  \frac{1}{\ln\lambda_i- \ln\lambda_j}. 
\nonumber 
\eeq
 The  form of $p_{ij}^\text{log}$ agrees with the formula for $ \tens{P}^\text{log} $ derived by  Xiao et al. \cite[eq. (41)]{Xiao97}.  Note that
\beq{-58}
\sgn p_{ij}^\text{log}  = \sgn p_{ij}^R = - \sgn  (p_{ij}^\text{log}-p_{ij}^R). 
\nonumber  
\eeq
The implications are twofold.  The first equalities indicate that  the spin induced by both 
${\bd \Omega}^\text{log}$ and by ${\bd \Omega}^R$ are 
 in the same sense, relative to the underlying spin $\bd W$.  The second equalities imply that the relative spin induced by ${\bd \Omega}^\text{log}$ is of smaller magnitude  than that of  ${\bd \Omega}^R$. 

The f-spin, which  is uniquely defined by the strain measure $f$, defines the skew matrix elements $p_{ij}^f$.  Consider the reverse problem: given some objective corotational rate defined by elements $p_{ij}$, is there a function $f$ such that   
$p_{ij}^f= p_{ij}$?    There is no such function for  the spins ${\bd \Omega}^R$, ${\bd \Omega}^E$ and ${\bd \Omega}^L$, as the reader can readily verify.  Obviously, $f = \ln$ for ${\bd \Omega} = {\bd \Omega}^\text{log}$, but it  remains an open question for general ${\bd \Omega}$ whether a strain measure function exists such that ${\bd \Omega}= {\bd \Omega}^f$.

\section{Eulerian conjugate  stress-strain pairs}\label{sec5}

\subsection{Arbitrary strain and corotational rate}

It was noted in the Introduction that the concept  of work-conjugate stress-strain pairs is  more complicated for Eulerian quantities  owing to the fact that the connection between the strain rate and the stretching tensor is not evident \emph{a priori}. 
  This issue was  resolved by Lehmann and Liang \cite{Lehmann93} who introduced the notion that the Eulerian pair ${\bd \tau}$ and ${\bd e}$ are defined to be conjugate if 
\beq{193}
\dot{w} = \tr \big( {\bd Q}{\bd \tau}{\bd Q}^t\, \dot{ \overline{ {\bd Q} {\bd e}{\bd Q}^t }} \big),  
\eeq
for some rotation ${\bd Q}$.  This clearly generalizes the Lagrangian work-conjugacy condition \rf{-10}, but it is necessary because  of the fact that Eulerian rates are not as restricted.   The definition \rf{193} is  equivalent to 
\beq{194}
\dot{w} = \tr ( {\bd \tau} \stackrel{\circ}{{\bd e}} ), 
\eeq
where $\stackrel{\circ}{{\bd e}} = \dot{\bd e}  + \{ {\bd e},  {\bd \Omega} \}$ and 
${\bd \Omega} = - {\bd Q}^t \dot{\bd Q}$.  Equation \rf{194} is taken as the starting point since it   depends only on the corotational  rate through the spin ${\bd \Omega}$, i.e. ${\bd Q}$ is not required. 

For a given strain measure ${\bd e} = f({\bd V})$ and corotational  rate ${\bd \Omega} = {\bd W} + \tens{P}{\bd D}$ 
the strain rate  $\stackrel{\circ}{{\bd e}} $ follows from Lemma \ref{-1}.  The stress ${\bd \tau}$ is therefore conjugate to 
${\bd e}$ if the following holds for all stretching tensors ${\bd D}$,
\beq{+45}
\tr \big( {\bd \tau} [\nabla f({\bd V})]\, \tens{Q}({\bd V}) \, D \big)
= \tr \big( {\bd \sigma} \, {\bd D}   \big).
\nonumber 
\eeq
The fourth order tensor $\nabla f({\bd V})$ is invertible for all strain measures.  The necessary and sufficient condition required to determine  $ {\bd \tau} $  is therefore that the
 fourth order tensor $\tens{Q}$ is invertible.   This requirement was obtained by Xiao et al. \cite{Xiao97} in a slightly different manner; basically, that the six elements $q_{ij}$ of eq. \rf{-+-} are all non-zero.   Hence,   $q_{ij}^{-1} $ are bounded, and $\tens{Q}^{-1}$ exists.  We refer the reader to Xiao et al. \cite{Xiao97} for further details.   
 
 In summary, the conjugate stress is 
 \beq{5-5}
 {\bd \tau} =  [\triangle f({\bd V})]  \tens{Q}^{-1} {\bd \sigma}, 
 \nonumber 
 \eeq
 where the order of  $[\triangle f({\bd V})] $ and $\tens{Q}^{-1}$ is arbitrary since they commute.  This is Theorem \ref{thm1}. 
 
 \subsection{Conjugate stress and the  f-rate}

 An alternative approach is suggested by eq. \rf{-45}. Let $\stackrel{\circ}{{\bd e}} = 
\tens{F} {\bd D}$, then the fourth order tensor $\tens{F}$ is by assumption invertible
and the conjugate stress is   simply ${\bd \tau} =   \tens{F}^{-1} {\bd \sigma}$. 
The tensor $\tens{F}$ can be  obtained directly in spectral form from eq. \rf{-45} and easily inverted, to give:  
\begin{lem}\label{lem9}
For arbitrary strain measure and rate the conjugate stress can be expressed 
\beq{-47}
{\bd \tau} =   {\bd \sigma}^f - 
\sum\limits_{\substack{i,j=1 \\ i\ne j}}^n  \frac1{1 +\big[(p_{ij} - p_{ij}^f ) \big( f(\lambda_i)- f(\lambda_j)\big) \big]^{-1}}
\, {\bd V}_i \boxtimes {\bd V}_j  \,
{\bd \sigma}. 
 \eeq
The conjugate stress satisfies 
 \beq{-08}
|{\bd \tau} -   {\bd \sigma}|^2 = |{\bd \tau} -     {\bd \sigma}^f|^2 + | {\bd \sigma}^f -   {\bd \sigma}|^2 , 
\eeq 
where the modified stress tensor $ {\bd \sigma}^f$ is  
\beq{06}
{\bd \sigma}^f = {\bd \sigma}
+ \sum\limits_{i=1}^n  \big[\frac{1}{
 \lambda_i f'(\lambda_i)}- 1
\big] 
{\bd V}_i \boxtimes {\bd V}_i  \, {\bd \sigma}.  
\nonumber 
\eeq
\end{lem}
The proof follows from eq. \rf{-47}
by analogy with the proof of  Lemma \ref{lem14}.
Hence,  $|{\bd \tau} -   {\bd \sigma}|^2 \ge |{\bd \sigma}^f -   {\bd \sigma}|^2$ 
with equality iff ${\bd \Omega} = {\bd \Omega}^f$, and we deduce:
\begin{lem}\label{lem16}
For every Eulerian strain measure $f( {\bd V})$ the corotational rate of the f-spin  ${\bd \Omega}^f$ minimizes the difference between the
conjugate stress 
 and the Cauchy stress.  The conjugate stress is then ${\bd \tau}= {\bd \sigma}^f$ and the minimal difference is 
\beq{085}
|{\bd \tau} -   {\bd \sigma}|^2 = \sum\limits_{i=1}^n \big[ 
\big( \frac1{\lambda_i f'(\lambda_i)}- 1\big) \tr ( {\bd V}_i{\bd \sigma})\big]^2.
\nonumber 
\eeq 
\end{lem}
This proves Theorem \rf{thm3}. 

In general $  {\bd \sigma}^f$ is not equal to the Cauchy stress for any strain measure, with the exception of  $f= \ln$, discussed below.  It is however, possible for  $ {\bd \sigma}^f$ and $ {\bd \sigma}$ to coincide under special circumstances:    if the three elements  $\tr ( {\bd V}_i{\bd \sigma})$ simultaneously vanish.  This is by definition a state of pure shear \cite{Norris06b}.  Hence, we have 
\begin{lem}\label{lem2}
If the Cauchy stress is a state of pure shear with $\diag {\bd \sigma} = 0$ in the principal axes of $\bd V$, then 
\beq{060}
 {\bd \sigma}^f   =  {\bd \sigma} . 
 \nonumber 
\eeq
The  stress conjugate to $f({\bd V})$ equals the Cauchy stress if the f-rate is used. 
\end{lem}
If the material is isotropic then the stress and strain share the same triad of principal axes. In that case $\diag {\bd \sigma} $ expressed in the  principal axes of $\bd V$ is simply the principle stresses, which vanishes only in the absence of stress.  Hence the circumstances under which Lemma \ref{lem2} applies cannot occur for isotropic materials.   If the material is not isotropic but we restrict attention to  linear anisotropic elasticity, then $\diag {\bd \sigma} $ expressed in the principal axes of strain $\bd e$ will vanish only if both stress and strain are zero.  This follows from the assumed positive definite property of  the strain energy, equal to $\frac12 \tr( {\bd \sigma}{\bd e})$.   In summary, the circumstances under which 
Lemma \ref{lem2} apply require nonlinear and anisotropic elasticity.     This does not eliminate its  possibility but it makes it difficult to envisage a situation when Lemma \ref{lem2} would occur. 

\subsection{Logarithmic rate}

The  logarithmic  rate, as noted before, is a special case of the f-rate.   We conclude by examining the conjugate stress for arbitrary strain measure using the  logarithmic rate.  Xiao et al.  \cite{Xiao97} 
showed that the logarithmic  rate  is the only one with the property of Lemma \ref{lem6}, i.e. among all strains and all rates, only $\ln {\bd V}$ and  ${\bd \Omega}^\text{log}$ correspond to the stretching tensor $\bd D$.   
This  fundamental result for $\ln {\bd V}$ is generalized to arbitrary Eulerian strain measure ${\bd e} = f({\bd V})$ by  
\beq{43}
\stackrel{\circ_\text{log} }{{\bd e}~~} 
= \big( \nabla f({\bd V})\big) \,  (\triangle \ln {\bd V} ) \, 
{\bd D}, 
\nonumber 
\eeq
which follows from Lemmas \ref{lem4} and \ref{lem6}.  Now require that the work-conjugacy identity $\tr ( {\bd \tau} \stackrel{\circ_\text{log} }{{\bd e}~~ })  = \tr ( {\bd \sigma}{\bd D})$   holds for all $\bd D$, and  use the invertibility of the fourth order tensors $\nabla f({\bd V})$  and $  \triangle \ln {\bd V}  $ plus  the property  that they commute.  This implies that 
the stress conjugate to the Eulerian strain ${\bd e} = f({\bd V})$ is 
\beq{-02}
 {\bd \tau}  = \big( \triangle f({\bd V})\big) \, (\nabla \ln {\bd V})
  \, {\bd \sigma} \qquad \text{for}\quad {\bd \Omega} = {\bd \Omega}^\text{log}.
 \nonumber 
 \eeq
It is straightforward to show that this can be expressed in spectral  form as 
 \beq{-027}
 {\bd \tau}  = {\bd \sigma}^f 
 + 
 \sum\limits_{ i,j=1 }^n \big( \frac{ \ln \lambda_i  - \ln \lambda_j }{f(\lambda_i) -  f(\lambda_j)}  - 1
 \big) {\bd V}_i \boxtimes {\bd V}_j \, {\bd \sigma} 
 \qquad \text{for}\quad {\bd \Omega} = {\bd \Omega}^\text{log}. 
 \eeq
This identity, although  valid only for the logarithmic rate, shows how the conjugate stress in that case is related to the modified stress ${\bd \sigma}^f$.  The latter depends upon the strain measure $f$, and is optimal in the sense of best for all possible strain rates.  Equation \rf{-027} shows that the logarithmic rate is not optimal since $\tau$ satisfies 
eq. \rf{-08} with both terms on the RHS of the latter non-zero.  However, when the strain measure $f$ reduces to $\ln$ then 
${\bd \sigma}^f  \rightarrow {\bd \sigma}$ and the sum in \rf{-027} vanishes.  This again shows the combined  properties of the Hencky strain and the logarithmic rate as being doubly optimal for all strain measures and spins.

\section{Conclusion}\label{sec6}

We have examined the implications of  work-conjugacy with emphasis on Eulerian stress-strain pairs.  There is, however,  remarkable similarity in the form of the dual conjugate stresses for both  Lagrangian and Eulerian strains.     The similarity is evident from the identical format of eqs. \rf{-4} and \rf{-12}, which involve fundamental stresses ${\bd T}^{(1)}$  and ${\bd \tau}^{(1)}$ defined by the strains 
$f({\bd U})$ and  $f({\bd V})$, respectively.  The Lagrangian stress ${\bd T}^{(1)}$ is called  Biot stress or Jaumann stress, but there does not appear to be a common term for its Eulerian counterpart ${\bd \tau}^{(1)}$.  

The major distinction between Lagrangian and  Eulerian work-conjugacy is that the latter requires the introduction of the  corotational  rate, which itself is quite arbitrary.  We have shown that  every permissible Eulerian strain measure  $f({\bd V})$  has associated with it a unique corotational  rate, the f-rate.  The conjugate stress obtained using the f-rate is optimal in the sense that it is the closest possible to the Cauchy stress ${\bd \sigma}$.   The optimal stress, ${\bd \sigma}^f$, is defined by $f$ and ${\bd \sigma}$ through eq. \rf{06}, and it reduces to the Cauchy stress if and only if $f=\ln$.  This reinforces the results of Xiao et al.  \cite{Xiao98c} for the the logarithmic rate and the Hencky strain, while generalizing the notion of the logarithmic rate to arbitrary strain functions through the strain dependent spin ${\bd \Omega}^f$. 

\section*{Acknowledgment}  
Advice from Ellis  Dill is gratefully appreciated. 
 
\appendix
\section{The spins ${ \Omega}^R $, ${ \Omega}^E $ and ${ \Omega}^L $}\label{appb}  

From the definition of ${\bd \Omega}^R$ in \rf{404}, and using ${\bd F}={\bd R}{\bd U}$, we have 
\beq{330}
{\bd L} = \dot{\bd F}{\bd F}^{-1} = {\bd \Omega}^R + {\bd R}\dot{\bd U}{\bd U}^{-1}{\bd R}^t.
\nonumber 
\eeq
The symmetric and skew parts of this relation yield \cite{Truesdell65}
\beq{302}
{\bd D}  =\frac12 {\bd R} ( \dot{\bd U}{\bd U}^{-1} + {\bd U}^{-1}\dot{\bd U} ) {\bd R}^t,
\qquad 
{\bd W}={\bd \Omega}^R + 
\frac12 {\bd R} ( \dot{\bd U}{\bd U}^{-1} - {\bd U}^{-1}\dot{\bd U} ) {\bd R}^t .
\eeq
Equation $\rf{302}_1$ may be solved for $\dot{\bd U}$ in the form given by 
eq. \rf{3051}.  Substituting $\dot{\bd U}$ in $\rf{302}_2$  gives 
\beq{306}
{\bd \Omega}^R  = {\bd W} + 
(  {\bd I}\boxtimes{\bd V} - {\bd V}  \boxtimes{\bd I}   )
(  {\bd I}\boxtimes{\bd V} +{\bd V}  \boxtimes{\bd I}   )^{-1} 
{\bd D}. \nonumber 
\eeq
 
Let ${\bd v}_i$, $i=1,\ldots , n\le 3$, be the principal axes of $\bd B$ and $\bd V$.  The twirl  ${\bd \Omega}^E $  defines the rate of rotation of this triad by  $\dot{\bd v}_i = {\bd  \Omega}^E{\bd v}_i$.  The rate of change of the eigentensors of $\bd B$ follows from ${\bd V}_i= {\bd v}_i\otimes {\bd v}_i$ as $\dot{\bd V}_i = \{{\bd  \Omega}^E, {\bd V}_i\}$.  Equation  $\rf{72}_2$ then gives
 \beq{+68}
 \dot{\bd B}  = \sum_{i=1}^n\dot{\beta}_i {\bd V}_i + \{{\bd  \Omega}^E, {\bd B}\},
 \eeq
  which can be considered as an equation for   
 ${\bd  \Omega}^E $, similar to eq. \rf{-55}.  The solution follows from eqs. \rf{-56} and \rf{4} as 
\bal{3606}
{\bd \Omega}^E  &= 
(  {\bd I}\boxtimes{\bd B} -{\bd B}  \boxtimes{\bd I}   )^*\, \big( \dot{\bd B} - \sum_{i=1}^n\dot{\beta}_i {\bd V}_i\big)
\\ \nonumber 
& = 
{\bd W} + 
(  {\bd I}\boxtimes{\bd B} -{\bd B}  \boxtimes{\bd I}   )^*
(  {\bd I}\boxtimes{\bd B} +{\bd B}  \boxtimes{\bd I}   ){\bd D}.
\end{align}
Hence ${ \bd \Omega}^E = {\bd W} + \tens{P}^E {\bd D}$ where $\tens{P}^E$ is given by \rf{-406b}.  The rate of change of the principal stretches are obtained by substituting ${\bd \Omega}^E$ into \rf{+68}, as
 \beq{+69}
 \sum_{i=1}^n\dot{\beta}_i {\bd V}_i 
 = \big[ \tens{I} - (  {\bd I}\boxtimes{\bd B} -{\bd B}  \boxtimes{\bd I}   )^*
 (  {\bd I}\boxtimes{\bd B} -{\bd B}  \boxtimes{\bd I}   )\big] 
 (  {\bd I}\boxtimes{\bd B} +{\bd B}  \boxtimes{\bd I}   ){\bd D}. 
 \nonumber 
 \eeq
Then using eqs. \rf{038} and \rf{67}, we obtain the well known result
 \beq{+70}
 \sum_{i=1}^n\dot{\beta}_i {\bd V}_i 
 =  2\sum_{i=1}^n {\beta}_i {\bd V}_i \boxtimes{\bd V}_i\, {\bd D}
 \qquad  
 \Leftrightarrow
 \qquad  \dot{\lambda}_i = {\lambda}_i\, \tr ( {\bd V}_i  {\bd D}). 
 \nonumber 
 \eeq
 
 The twirl  $\tilde{\bd  \Omega}^L $ defines  the rate of rotation of the Lagrangian principal axes 
 ${\bd u}_i$, $i=1,\ldots , n$ as $\dot{\bd u}_i = \tilde{\bd  \Omega}^L{\bd u}_i$. Hence, 
 $\dot{\bd U}_i = \{ \tilde{\bd  \Omega}^L, {\bd U}_i\} $ where ${\bd U}_i = {\bd u}_i\otimes {\bd u}_i$
are the eigentensors of  ${\bd U}$.  Taking the rate of change of the identity  ${\bd V}_i = ({\bd R}\boxtimes {\bd R}) {\bd U}_i$ linking the Eulerian and Lagrangian 
eigentensors,   gives 
 \beq{-98}
  \dot{\bd U}_i = ({\bd R}\boxtimes {\bd R})^{-1} \, \{ {\bd  \Omega}^E- {\bd  \Omega}^R, {\bd V}_i\}. 
  \nonumber 
  \eeq
The Lagrangian twirl is therefore 
  \beq{-78}
 \tilde{\bd  \Omega}^L = ({\bd R}\boxtimes {\bd R})^{-1} \,( {\bd  \Omega}^E- {\bd  \Omega}^R )
 = (  {\bd I} \boxtimes{\bd U}^2 -{\bd U}^2\boxtimes{\bd I} )^*
\, 2   ({\bd U} \boxtimes{\bd U} ) \, ({\bd R}\boxtimes {\bd R})^{-1} \, {\bd D}. 
\nonumber 
  \eeq
This is related to the spin ${\bd  \Omega}^L = {\bd W} + \tens{P}^L {\bd D}$ defined via $\tens{P}^L$ of eq. \rf{-406c}  by 
$\tilde{\bd  \Omega}^L = ({\bd R}\boxtimes {\bd R})^{-1} ({\bd  \Omega}^L - {\bd W}) $.  


\begin{thebibliography}{10}

\bibitem{Curnier91}
A.~Curnier and L.~Rakotomanana.
\newblock Generalized strain and stress measures: critical survey and new
  results.
\newblock {\em Engineering Transactions (Polish Academy of Sciences)},
  39(3-4):461--538, 1991.

\bibitem{Dill06}
E.~H. Dill.
\newblock {\em Continuum Mechanics: Elasticity, Plasticity, Viscoelasticity}.
\newblock CRC Press, 2006.

\bibitem{Dui06}
G-S Dui.
\newblock Some basis-free formulae for the time rate and conjugate stress of
  logarithmic strain tensor.
\newblock {\em J. Elasticity}, 83(2):113--151, May 2006.

\bibitem{Dui07}
G-S Dui, Z.~Wang, and Q.~Ren.
\newblock Explicit formulations of tangent stiffness tensors for isotropic
  materials.
\newblock {\em Int. J. Num. Meth. Engin.}, 69(4):665--675, 2007.

\bibitem{Fitz80}
E.~J. Fitzgerald.
\newblock A tensorial {H}encky measure of strain and strain rate for finite
  deformations.
\newblock {\em J. Appl. Phys.}, 51(10):5111--5115, 1980.

\bibitem{Hill78}
R.~Hill.
\newblock {Aspects of invariance in solid mechanics}.
\newblock {\em Adv. Appl. Mech.}, 18:1--75, 1978.

\bibitem{Hoger87}
A~Hoger.
\newblock The stress conjugate to logarithmic strain.
\newblock {\em Int. J. Solids Struct.}, 23(12):1645--–1656, 1987.

\bibitem{Lehmann93}
T.~Lehmann and Haoyun Liang.
\newblock {The stress conjugate to logarithmic strain In V}.
\newblock {\em Z. Angew. Math. Mech.}, 73(12):357--363, 1993.

\bibitem{Lehmann91}
T.~H. Lehmann and Z.~Guo.
\newblock {The conjugacy between Cauchy stress and logarithm of the left
  stretch tensor}.
\newblock {\em Eur. J. Mech. A}, 10(4):395--404, 1991.

\bibitem{Macvean68}
D.~B. MacVean.
\newblock {The elementary work in a continuum and the correlation of stress and
  strain tensors}.
\newblock {\em Z. Angew. Math. Phys.}, 19:157--185, 1968.

\bibitem{Norris06b}
A.~N. Norris.
\newblock Pure shear axes and elastic strain energy.
\newblock {\em Q. J. Mech. Appl. Math.}, 59:551--562, 2006.

\bibitem{Norris07b}
A.~N. Norris.
\newblock Higher derivatives and the inverse derivative of a tensor-valued
  function of a tensor.
\newblock {\em Q. Appl. Math.}, (accepted):1--1, 2007.

\bibitem{Ogden84}
R.~W. Ogden.
\newblock {\em Non-Linear Elastic Deformations}.
\newblock Ellis Horwood, 1984.

\bibitem{Rosati00}
L.~Rosati.
\newblock A novel approach to the solution of the tensor equation {AX+XA=H}.
\newblock {\em Int. J. Solids Struct.}, 37(25):3457--3477, June 2000.

\bibitem{Scheidler91}
M.~Scheidler.
\newblock {Time rates of generalized strain tensors. Part I: Component
  formulas}.
\newblock {\em Mech. Materials}, 11:199--210, 1991.

\bibitem{Truesdell65}
C.~Truesdell and W.~Noll.
\newblock {\em The non-linear field theories of mechanics}, volume III of {\em
  Encyclopedia of Physics}.
\newblock Springer-Verlag, Berlin, 1965.

\bibitem{Xiao95}
H.~Xiao.
\newblock Invariant characteristic representations for classical and micropolar
  anisotropic elasticity tensors.
\newblock {\em J. Elasticity}, 40:239 -- 265, 1995.

\bibitem{Xiao06}
H.~Xiao, O.~Bruhns, and A.~Meyers.
\newblock Elastoplasticity beyond small deformations.
\newblock {\em Acta Mech.}, 182(1):31--111, March 2006.

\bibitem{Xiao97}
H.~Xiao, O.~T. Bruhns, and A.~Meyers.
\newblock Logarithmic strain, logarithmic spin and logarithmic rate.
\newblock {\em Acta Mech.}, 124(1):89--105, March 1997.

\bibitem{Xiao98c}
H.~Xiao, O.~T. Bruhns, and A.~Meyers.
\newblock Objective corotational rates and unified work-conjugacy relation
  between {E}ulerian and {L}agrangean strain and stress measures.
\newblock {\em Arch. Mech.}, 50(6):1015--1045, 1998.

\bibitem{Xiao98d}
H.~Xiao, O.~T. Bruhns, and A.~Meyers.
\newblock On objective corotational rates and their defining spin tensors.
\newblock {\em Int. J. Solids Struct.}, 35(30):4001--4014, October 1998.

\bibitem{Xiao98}
H.~Xiao, O.~T. Bruhns, and A.~Meyers.
\newblock Strain rates and material spins.
\newblock {\em J. Elasticity}, 52(1):1--41, July 1998.

\end{thebibliography}

\end{document}